# A Comprehensive Review of Technologies Used for Screening, Assessment, and Rehabilitation of Autism Spectrum Disorder

Shadan Golestan, Pegah Soleiman, Hadi Moradi

**Abstract**—Autism Spectrum Disorder (ASD) is an umbrella term for a wide range of developmental disorders. For the past two decades, researchers proposed the use of various technologies in order to tackle specific symptoms of the disorder. Although there exist many literature reviews about screening, assessment, and rehabilitation of ASD, no comprehensive survey of types of technologies in all defined symptoms of ASD has been presented. Therefore, in this paper a comprehensive survey of previous studies has been presented in which the studies are classified into three main categories, and several sub-categories, and three main technologies. An analysis of the number of studies in each category and sub-category is given to help researchers decide on areas which need further investigation. The analysis show that the majority of studies fall into the software-based systems technology category. Finally, a brief review of studies in each category of ASD is presented for each type of technology. As a result, this paper also helps researchers to obtain an overview of the typical methods of using a specific technology in ASD screening, assessment, and rehabilitation.

━━━━━━━━━━ ◆ ━━━━━━━━━━

## 1 Introduction

ACCORDING to the 5th edition of Diagnostic and Statistical Manual of Mental Disorders (DSM-V) [1], Autism Spectrum Disorder (ASD) is a neurodevelopmental disorder with irregularities that adversely affect developmental process of children. Children with autism, usually encounter wide range of mental and developmental disorders. They usually face difficulties in social and communication skills and restricted or repetitive behaviors [1], [2]. Unfortunately, in recent years the number of children who diagnosed with autism increased rapidly [3], [4]; such that 1 in 68 children in the United States is screened with autism [2].

The rehabilitation and assessment of these children is very difficult due to many factors, e.g. the difficulty in communicating with these children or their wide range of symptoms. Therefore, it is usually necessary to apply distinct methods for every individual with autism [2], [5]. Hence, the rehabilitation and assessment tasks are usually time consuming and put burden on caregivers or parents [6]. On the other hand, the screening, especially early screening at early years of a child, is crucial for successful rehabilitation.

Beside the above issues in dealing with children with autism, they are generally very interested in technological devices such as computers, smartphones, tablets, and robots [7]. This interest lies on the fact that they can predict outcomes of their actions. In contrast, they usually face difficulties in interaction[7] in the real world, which could be confusing to them and the outcome of their actions are usually unpredictable [7]. Moreover, technological devices are widely available which may reduce the cost of screening, assessment, and rehabilitation tasks. Therefore, utilizing them in the aforementioned tasks has been increased recently [8]–[12]. many related studies has been searched by using specific keywords[1] (Fig. 1) in order to demonstrate the growing trend of using technology in these tasks. The figure shows that rapid increase specially after 2004. The increase in the funding for autism research in recent years can be a reason for the increase in the research after 2010.

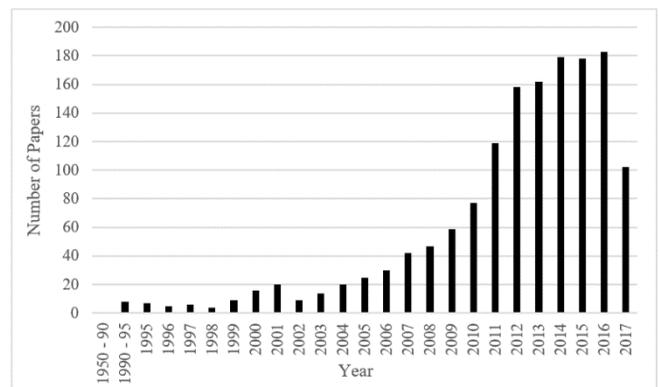

Fig. 1. Total number of papers that presented a technology-based diagnosis, intervention or assessment for autistic individuals in each year. A software called "Publish or Perish" is used to submit queries over titles of papers exist in Google Scholar database. The keywords were "autism" or "ASD" and any of the words "computer", "mobile", "smartphone", "tablet", "sensor", "robot", "robotics", "device", "software", iPad", "technology", "multimedia" and "game".

- *Shadan Golestan is with the School of ECE, University of Tehran, Tehran, Iran. E-mail: shgolestan@ut.ac.ir.*
- *Pegah Soleiman is with the School of Computer Science, University of Tehran, Tehran, Iran. E-mail: pgh.soleiman@ut.ac.ir.*
- *Hadi Moradi is with the School of Computer Science, University of Tehran, Tehran, Iran. He is also with Intelligent Systems Research Institute, SKKU, Suwon, South Korea. E-mail: moradih@ut.ac.ir.*

It should be mentioned that the data for 2017 is not accurate since many articles have not been posted online yet and we should wait for a while to have a more accurate number for 2017.

It is worth to say that there are surveys and literature reviews that focused on available assistive technologies for children with autism. Table 1 shows a list of several surveys categorized based on the type of technology used and based on the type of problem it is used for. The table shows that none of these literature reviews has provided a comprehensive categorization in neither both parts. Therefore, in this paper a comprehensive survey of available studies that used technology in order to tackle a difficulty of children with autism has been presented. The information provided in this survey may be used to determine possible use of technologies suitable for screening, assessment, and rehabilitation tasks. Also, potential and un-attempted areas for further research can be determined. Finally, the brief reviews of selected studies help the newcomers to this field to better understand the methodology and proper steps needed to step into this multidisciplinary field of research.

The rest of the paper is organized as follows. First of all, a distribution of types of utilized technologies in screening, assessment, and rehabilitation of children with autism will be presented. Then, it will be discussed that how DSM-V defines diagnostic ASD symptoms which are used to categorize the previous studies. Thereafter, the statistics of previous studies based on the different categories will be provided. Then, a discussion among tackled and unattempted areas in autism rehabilitation will be made. Finally, in each category, some prominent studies will be reviewed.

## 2 METHOD

### 2.1 Types of Technologies in Dealing with Autism

By reviewing previous studies, the technologies used in screening, assessment, and rehabilitation of children with autism can be divided into the following types:

**Software-based**: this type of systems is software oriented. Also, available hardware, such as smartphones or game consoles, are used without any focus on the hardware aspects.

**Hardware-based:** in contrast to the software-based systems, this type includes systems that focus mostly on designing or using a specific hardware. It should be mentioned that these devices are divided into two types since the majority of the studies either use: a) the available *Robots* or b) designed or used mechatronic devices, which they are called *Dedicated Devices* in this paper. The *Robots* type is referred to robotic devices that are either humanoid, such an NAO [13], look like living creatures, such as Keepon [14] or interactive mechatronics like Queball [15]. The *Dedicated Devices* are developed to collect data or interact with children such as the E4 wrist band [16] or the Intelligent Toy Car [17].

Fig. 2 shows the distribution of 212 reviewed papers in the mentioned technology types. As it can be inferred from Fig. 2 that software-based approach is obviously more popular than the hardware-based approach. The main reasons could be in the ease of design, development, use, and upgrade of a software-based system. Actually, the mentioned factors made the software-based systems more affordable and widely available for testing and use.

TABLE 1
PREVIOUS LITERATURE REVIEWS IN UTILIZING A TYPE OF TECHNOLOGY FOR REHABILITATION OF ASD CHILDREN IN SPECIFIC CRITERIA.

| Authors | Year | Focused type of technology ||||| Screening, Assessment, and Rehabilitation areas related to children with autism ||||||
|---|---|---|---|---|---|---|---|---|---|---|---|---|
| | | Mobile Applications* | Computer Games | Computer Software | Dedicated Devices** | Robots | Social skills | Communication skills | Learning skills | Visual motor skills | Sensory Integration | Daily living skills |
| Reed et al. [95] | 2011 | | | ✓ | ✓ | ✓ | ✓ | | | | | |
| Wainer et al. [96] | 2011 | | | | | ✓ | ✓ | ✓ | | | | |
| Ramdoss et al. [97] | 2011 | | | ✓ | | | | ✓ | | | | |
| Chen [98] | 2012 | | | | ✓ | | ✓ | ✓ | | | | |
| Noor et al. [99] | 2012 | ✓ | | | | | ✓ | ✓ | ✓ | ✓ | ✓ | |
| Shane et al. [100] | 2012 | | | ✓ | ✓ | | ✓ | ✓ | | | | |
| Scassellati et al. [101] | 2012 | | | | | ✓ | ✓ | ✓ | | | | |
| Ploog et al. [9] | 2013 | | | ✓ | | | ✓ | ✓ | | | | |
| Cabibihan et al. [102] | 2013 | | | | | ✓ | ✓ | ✓ | | | | |
| Knight et al. [103] | 2013 | ✓ | ✓ | | | | | | ✓ | | | |
| Ayres et al. [104] | 2013 | ✓ | | | | | ✓ | | ✓ | | | |
| Lang et al. [11] | 2014 | | | ✓ | ✓ | | ✓ | ✓ | | | | ✓ |
| Alzrayer et al. [105] | 2014 | ✓ | | | | | | ✓ | | | | |
| Lorah et al. [106] | 2014 | ✓ | | | | | ✓ | ✓ | | | | |
| Coeckelbergh et al. [107] | 2015 | | | | | ✓ | ✓ | ✓ | | | | |
| Jagdev et al. [108] | 2015 | | | | | ✓ | ✓ | | | | ✓ | |
| Brown et al. [109] | 2015 | | ✓ | ✓ | | | | | ✓ | | | |
| Ng et al. [110] | 2015 | | | ✓ | | | ✓ | | | | | ✓ |
| Zaman et al. [111] | 2015 | ✓ | | | | | ✓ | ✓ | ✓ | | | |
| McCoy et al. [112] | 2016 | | ✓ | ✓ | | | ✓ | ✓ | | | | |

\* This category refers to smartphone-based or tablet-based applications.
\*\* Multi-touch interfaces, video and sound recorders, tactile systems et cetera are placed in this category.

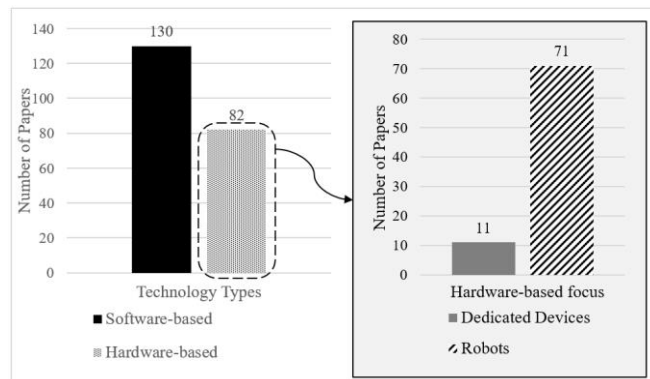

Fig. 2. Distribution of software-based and hardware-based technologies among the 212 reviewed papers.

## 2.2 Three Areas of Interest in ASD

Based on DSM-V [1], there are many prevalent symptoms in children with autism. As DSM-V suggests, these symptoms can be divided into two main categories, difficulties in social and communication skills and restricted or repetitive behaviors. Thereby, the studies can be classified into these two categories. Furthermore, our study shows that there are many studies that proposed systems for teaching daily life activities or learn academic materials [18]–[24]. Consequently, Adversities in Learning category has been added as the third category in autism screening, assessment, and rehabilitation. Fig. 3 depicts these categories and their sub-categories which are used to organize the related work.

Fig. 4 shows the number of studies reported in each of the main categories of Fig. 3 It can be concluded from Fig. 4 that most of the previous studies focused on the "Difficulties in Social and Communication Skills" of children with autism. It also shows that the software-based research is the most widely used approach. As mentioned earlier, here also the reason can be due to its ease of design, development, and test. Moreover, using robots are the second widely used approach since they are attractive for children with autism, fairly available to researchers, and easier to use than dedicated devices.

There exist few studies related to "Restricted or Repetitive Behaviors" of children with autism. The reason seems to be due to their less importance in the developmental process of children with autism. Also, it should be noted that the numbers of studies, using different technologies to deal with the restricted or repetitive behaviors, are fairly close. This could be due to the smaller impact of these behaviors in the life of children with autism.

Finally, utilizing software systems seem to be promising when studies try to improve "Adversities in Learning" of children with autism. Actually, no previous research using dedicated devices or robots has been found in this category. This is because of the fact that through software-based systems it is possible to offer mobile and instructive solutions to this difficulty.

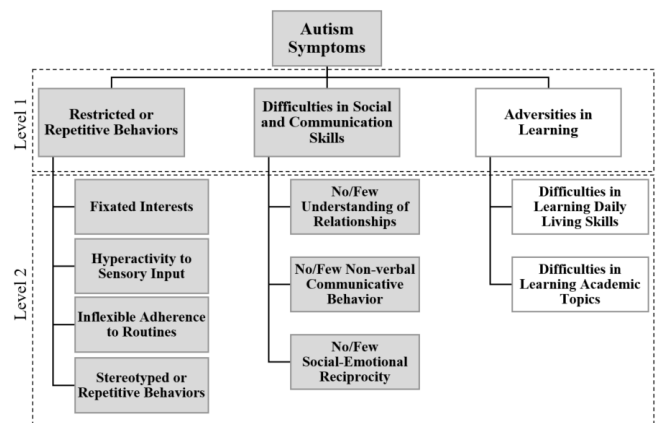

Fig. 3. The two categories obtained from DSM-V diagnostic criteria (Grey branches) and the Adversities in Learning category (white branch).

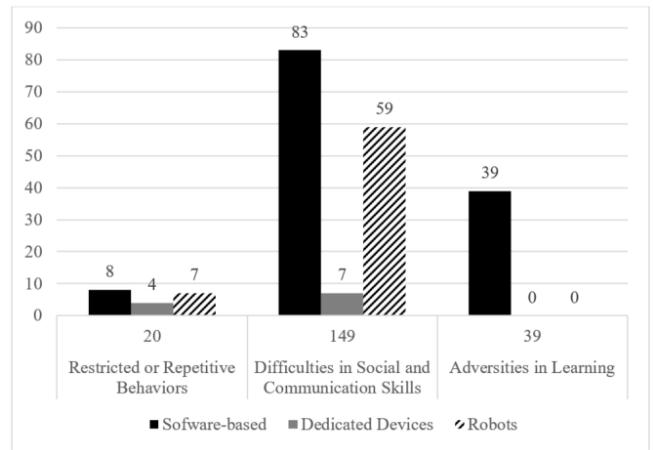

Fig. 4. The distribution of the studies in three categories related to autism screening, assessment, and rehabilitation.

A more detailed analysis of the studies in the sub-categories of Fig. 3, is shown in Fig. 5. As depicted in Fig. 5, there are no previous attempts in order to tackle "Fixated Interests" of children with autism. Also no studies based on dedicated devices has been found in "Hyperactivity to Sensory Input" sub-category. This is also true for "Inflexible Adherence to Routines" sub-category.

The figure shows that there are almost equal number of studies using software-based approaches and dedicated devices to deal with "Stereotyped or Repetitive Behaviors" of these children.

In the "No/Few Understanding of Relationships" sub-category, the software-based approaches are almost two times more than the hardware-based, i.e. robots and dedicated devices combined. However, the differences between software-based approaches and robots are not significant in "No/Few Non-verbal Communicative behaviors" and "No/Few Social-Emotional Reciprocity" sub-categories. To get more in depth analysis of the distribution of the approaches used in the "Difficulties in Social and Communication Skills", further sub-categories, called symptoms, using DSM-V have been determined (Table 2). Fig. 6 shows the distribution of studies in the symptoms of Table 2 that are specified in the "Difficulties in Social and Communication Skills".

As a general observation, software-based systems are noticeably dominant approaches among the others in "No/Few Social Interactions", "No/Few Initiation of Social Interaction", and "No/Few Share of Emotion". The reason behind this could be the fact that software-based systems often act as assistive mobile tools for children with autism in social interactions. On the other hand, robots have been more exploited in "No/Few Understanding of Joint Attention" and "No/Few Imitation of Other's Behavior". The reason is that interactions based on joint attention and imitation skills are surely more effective in 3D environments.

A generalization of the above observation suggests that the sub-categories that involves embodiment, such as understanding joint attention or gesture, can be better performed by robots. However, due to current limitations in the availability and capabilities of robots, they may have not been fully employed in the studies. On the other hand, the ease of software development, maintenance, and upgrade, beside it's lower development cost, made it the favorite technology in most of the studies. It is expected to see more use of robots in the future, since they become more sophisticated and more widely available, because they can be used in the majority of the categories and sub-categories.

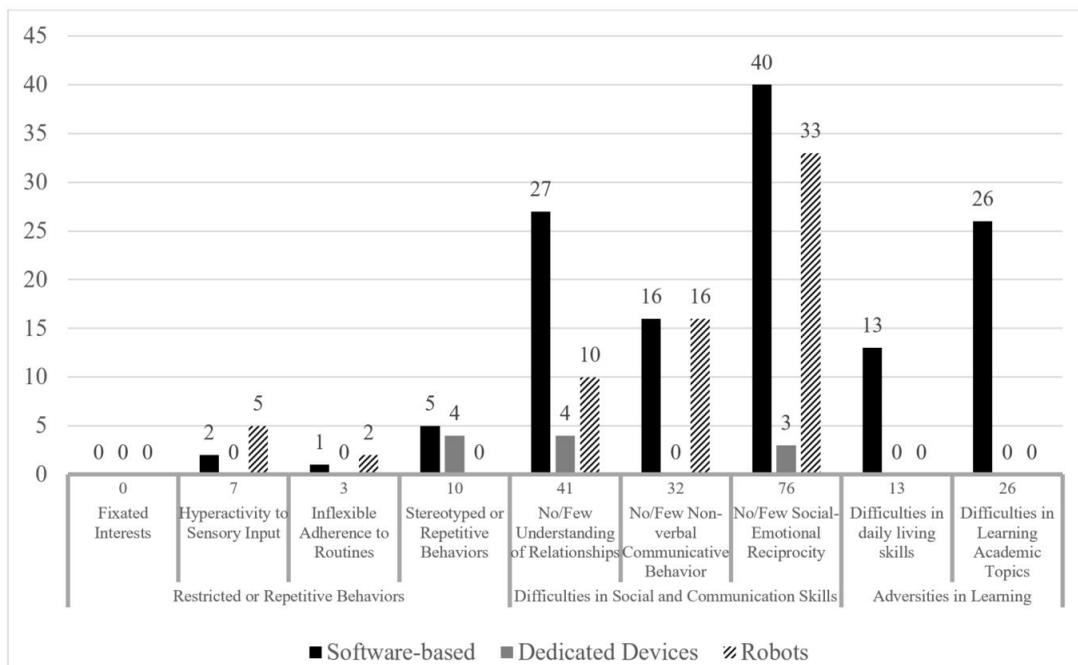

Fig. 5. The distribution of the studies in the sub-categories related to autism screening, assessment, and rehabilitation.

In the following, bold trends of studies in using the three types of technologies in each category are discussed. Therefore, typical methods of utilizing the technologies in each category will be revealed. Then, a brief review of studies in each category at the sub-category is provided.
For papers accepted for publication, it is essential that the electronic version of the manuscript and artwork match the hardcopy exactly! The quality and accuracy of the content of the electronic material submitted is crucial since the content is not recreated, but rather converted into the final published version.

## 3 BRIEF REVIEW OF STUDIES

### 3.1 Difficulties in Social and Communication Skills Category

As discussed earlier, there are numerous approaches used software-based systems in this category. In the following, a few software-based studies in this category have been mentioned. Many of the software-based studies focused on designing a system to teach emotions via virtual characters [25]–[32]. For instance, Fergus et al. [33] utilized the WoodyTM character from the Toy Story animation for encouraging children with autism to understand emotional facial expression.

Furthermore, a few software-based systems focused on developing applications to practice eye/gaze tracking [34], [35] and face recognition [36] for these children. In detail, Korhonen et al. [37] presented a learning game consists of a video game, a handheld computer, a Microsoft Kinect®, a projector, and SMI® eye-tracking glasses. The game includes a virtual character that provides either eye/gaze direction, finger pointing, or dotted arrow cues (or a combination of those).

Moreover, there are software-based approaches for improving social skills of children with autism that use specific collaborative play frameworks [38]–[41]. These frameworks involve number of children to play a multiplayer game together. In particular, Piper et al. [42] designed and developed a tabletop computer game which is best suited for collaboration of four children with autism.

TABLE 2
THE SYMPTOMS OF EACH SUB-CATEGORY RELATED TO "DIFFICULTIES IN SOCIAL AND COMMUNICATION SKILLS" SECTION.

| Main Category | Sub-categories | Symptoms |
|---|---|---|
| Difficulties in Social and Communication Skills | No/Few Understanding of Relationships | Inappropriate Approaches That Seem Aggressive or Disruptive |
| | | No/Few Social Interactions |
| | | Passivity |
| | No/Few Non-verbal Communicative misbehaviors | No/Few Eye Contact |
| | | No/Few Understanding of Body Orientation |
| | | No/Few Understanding of Gestures |
| | | No/Few Understanding of Joint Attention |
| | | No/Few Understanding of Speech Intonation |
| | No/Few Social-Emotional Reciprocity | No/Few Initiation of Social Interaction |
| | | No/Few Share of Emotion |
| | | No/Few Imitation of Other's Behavior |

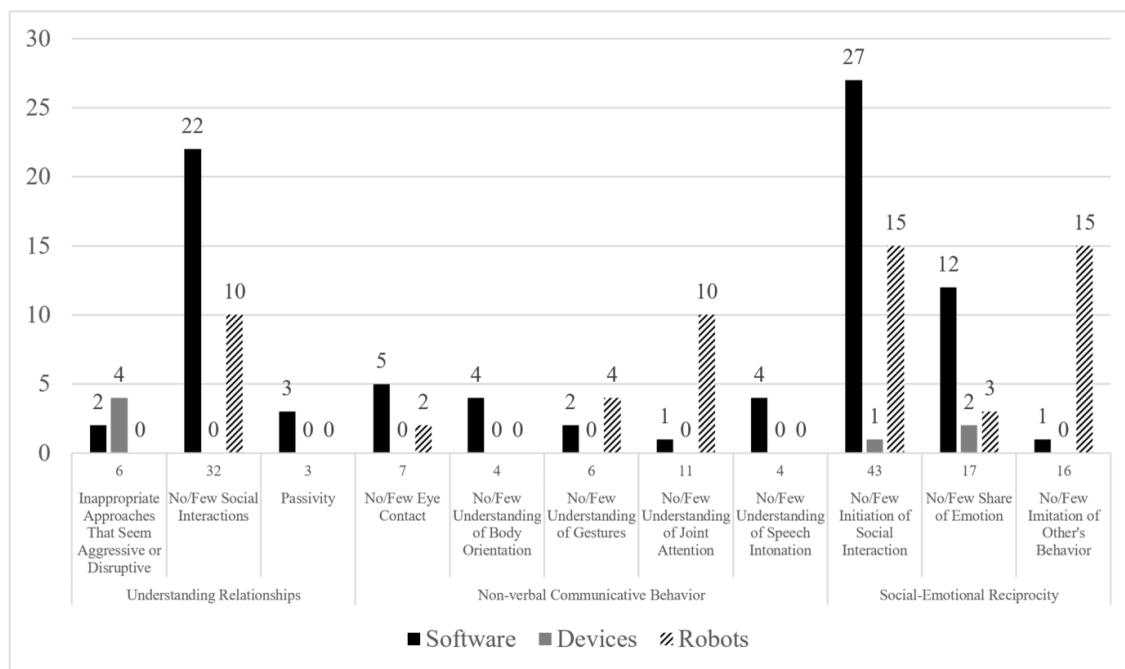

Fig. 6. Distribution of studies in symptoms of "Social Communication and Social Interaction" section.

In addition, many other software-based approaches proposed story-telling games in order to increase the narrative and imaginative skills of children with autism [43]–[46]. Namely, Pennington et al. [47] proposed a computer software that assists the children to complete a story.

In comparison to software-based systems, devices have been barely used in improving social and communications skills of children with autism. A few studies are available such as the work presented by C. Liu et al. [48] which shows that using ECG can help in recognizing the emotions of children with autism. Also, B. Knapp et al. [49] show how EMG can be used for emotion regulation in children with autism. Moreover, Özcan et al. [48] believe that using a wearable device, called "+Me", could motivate children with autism to communicate and interact more with others. Finally, there studies suggest that using tangible user interfaces, which involve physical and tangible objects in a cooperative game, could increase social skills of children with autism [50], [51].

In order to show the benefits of using robots in Difficulties in Social and Communication Skills category, we can give examples of robots that are designed to improve joint attention and imitation skills of children with autism [50]–[56]. In particular, Fujimoto et al. [56] studied suitable methods for mimicking human motion by a humanoid robot in order to design sessions for improving imitation skills of children with autism.

Also, many studies suggest that using robots as an assistance in therapy sessions could increase the therapy effects on children with autism [57]–[61]. As an example of using robots as assistance to therapists, Vanderborght et al. [62] studied the use of a huggable robot, called Probo, as a social storytelling agent for children with ASD. Probo is able to express attention and emotions via its gaze and facial expressions.

What comes next is a brief review of a few studies that used a specific technology in Difficulties in Social and Communication Skills sub-categories.

### 3.1.1 No/Few Social-Emotional Reciprocity sub-category

Software sample:

*MOSOCO: A Mobile Assistive Tool to Support Children with Autism Practicing Social Skills in Real-Life Situations* [32]. This application uses augmented reality and visual supports of a validated curriculum, from the Social Compass, to help children with autism practice social skills in real-life situations. The original Social Compass curriculum is limited to classrooms and currently have limited support for mobile platforms. Thus, children with autism usually face difficulties when they want to generalize what they learn in classrooms and therapy sessions to real-life situations. MOSOCO facilitates practicing social skills in real-life situations, guiding students through the six fundamental social skills of the Social Compass curriculum which are: eye contact, space and proximity, start an interaction, asking questions, sharing interests and finish an interaction. The use of this application, increased the interaction between students alongside they have practiced basic steps of a proper interaction with others, how to identify and avoid missteps and how to help their partners interact. One of the key advantages of this systems is that it can be used outside of classrooms. In addition, the system provides students confidence in interaction with others and improve their skills by repetition.

Devices Sample:

*Feasibility Study in Development of a Wearable Device to Enable Emotion Regulation in Children with Autism Spectrum Disorder* [63]. In this research the proposed device uses physiological data in order to estimate level of stress and tries to regulate it by playing some audio through a feedback loop system. Their results seem to be very promising. In detail, the proposed system was able to determine whether the

situation is relaxed or stressed with 83% accuracy. Also it was mostly successful in capturing children's attention by playing audio.

### Robots Sample:

*Autonomous Robot-mediated Imitation Learning for Children with Autism* [64]. The researchers proposed an autonomous and adaptive robotic system suitable for improving imitation skills of autistic children in a co-robotic intervention setting. The system evaluates and gives feedback in real-time basis. It has been stated that the co-robotic intervention has better performance in comparison to human-oriented sessions. They used a humanoid robot to show a desired gesture to the child and used a Microsoft Kinect in order to capture the imitated gesture by the child to measure the accuracy of the imitation. For evaluating the results, they provided two separate rooms, one for pre and post assessments and one for intervention. Their results show that children with autism spent more time interacting with robots in comparison to interaction with human.

### 3.1.2 No/Few Non-verbal Communicative Behavior sub-category

### Software sample:

*VocSyl: Visualizing Syllable Production for Children with ASD and Speech Delays* [65]. According to Heilpern et al. [65], the majority of autistic children are visual learners. Hence, they designed and developed a software that visualizes sound. Through this system, they tried to effectively educate children who suffer speech difficulties such as children with autism. Their software visualizes the syllable of words in order to make a visual difference between different syllables. Therefore, parents, caregivers, or therapists can easily visualize the correct syllable of a desired word and have children to pronounce that word in a way that matches their visualized syllable.

### Robots Sample:

*How children with autism spectrum disorder behave and explore the 4-dimensional (spatial 3D + time) environment during a joint attention induction task with a robot* [52]. The researchers studied the way that children with autism behave with their 4 dimensional environment when interacting with a Nao robot or a human when practicing joint attention. One of the main observations of their research was that autistic individuals had less stable body orientation in compare to typically developed children. This might be one of the reasons that these children face difficulties in social interaction skills.

### 3.1.3 No/Few Understanding of Relationships sub-category

### Software sample:

*Multitouch Tablet Applications and Activities to Enhance the Social Skills of Children with Autism Spectrum Disorders* [40]. Hourcade et al. [40] focused on the impact of computer-based approaches in increasing the social skills of children diagnosed with Autism Spectrum Disorders (ASD). They believe that despite a great enhancement in diagnosing ASD in early ages, much of their needs in childhood remains unanswered which will cause social interaction difficulties in adulthood. Therefore, by using multitouch tablets and surfaces, they presented an application to improve the social skills of these children, with concentration on collaboration, coordination, creativity, compromising, and understanding emotions. They designed their applications very user friendly and fault tolerant to decrease the frustration and open-ended to be flexible enough. In the following, a brief description of their designed applications are mentioned:

- *Drawing*. Through drawing activities, they provide children with ASD to express their feelings quickly through drawing. Moreover, they used this application to enhance the social skills of these children by engaging them in storytelling activities. In such a scenario, a group of children collaborate with each other in telling a story by drawing on a device one by one. They believe that this approach made a huge improvement in the children's social skills.
- *Music Authoring.* This application makes it possible for children to compose a desired music by playing a visual instrument. This will allow children to share their composed music with others. Similar to the drawing application, here also researchers tried to gather a group of children to collaborate with each other and author a piece of music.
- *Untangle.* This application is actually a visual puzzle in which there are a number of circles connected to two other circles by two lines. To solve the puzzle, children should move the position of the circles in a way that there will be no line intersecting another line. The outcomes of this application, when played in group, is to improve social interaction skills such as communication alongside visuospatial thinking of individuals. The design of this puzzle encourages children to collaborate and coordinate to solve it.
- *Photogoo.* This application allows children to distort a picture by dragging their fingers on it. This application is designed such that by distorting a picture, which shows a face, low-functioning ASD children can easily learn and practice emotions.

### Devices Sample:

*Social Benefits of a Tangible User Interface for Children with Autism Spectrum Conditions* [66]. Farr et al. [66] studied the impact of assembling a toy, named Topobo which is a programmable Lego kit, in improvement of social interaction in children suffering from autism. Actually, LeGoff [67] stated that playing with a basic Lego kit will increase the social skills of children with autism. Furthermore, Farr et al. [66] showed that children play more with Topobo than a basic Lego kit when they are not alone.

### Robots Sample:

*Collaborating with Kaspar: Using an Autonomous Humanoid Robot to Foster Cooperative Dyadic Play among Children with Autism* [61]. In this paper, the effect of interacting with a humanoid robot, i.e. Kaspar, is studied. The researchers were interested to see how interacting with the robot can affect positively on interaction between children with autism and adults. Their promising results show that the children interact better with their parents after an interaction session with the humanoid robot.

## 3.2 Repetitive and Restricted Behaviors Category

In this category, there exist few software-based studies that tried to address the difficulty of hyperactivity to sensory input in children with autism [68], [69]. In particular, Chuah et al. [69] used various smartphone sensors to recognize several repetitive behaviors of children with autism, i.e. hand waving, jumping, foot tapping, walking, and sitting. They also analyzed the sounds in their environment to predict the situation that gave rise to a specific behavior.

There are also studies that proposed vision-based behavior recognition systems that facilitate the recognition and assessment of abnormal behaviors. Particularly, recognizing such stereotyped and repetitive behaviors, will be done more easily and effectively [70], [71]. As a result, the screening, assessment, and rehabilitation tasks will be done with more accuracy.

The use of Dedicated-Devices in this category is focused on Stereotyped and Repetitive Behavior sub-category [72]–[74]. For instance, Albinali et al. [72] have proposed an activity recognition system which is capable of recognizing stereotyped hand-flapping and body-rocking behaviors. They have used three wearable accelerometer sensors in both hand wrists and torso to collect discriminative data.

One of the uses of Robots in this category is to address the difficulty of hyperactivity to sensory input in children with autism [75], [76]. For instance, Robins et al. [75] investigated that embedding tactile sensors in a humanoid robot, i.e. KASPAR can help children with autism to improve their tactile interaction.

What comes next is a brief review of studies that used a particular technology to tackle the Stereotyped and Repetitive Behavior sub-categories.

### 3.2.1 Hyperactivity to Sensory Input Sub-category

Software sample:

*Smartphone based autism social alert system* [69]. The proposed software, is a smartphone application capable of obtaining data from the accelerometer sensors of smartphones or from wearable sensors in order to recognize behaviors like foot tapping, waving, jumping, walking, and sitting. The gathered data will be sent to a server and a classification algorithm on the server, will detects the abnormal behaviors from the data. The system is also capable of determining the cause of repetitive behavior of these children, e.g. sudden or loud sounds in their environments.

Devices Sample:

*Investigating tactile event recognition in child-robot interaction for use in autism therapy* [77]. This study focuses on tactile feedback and its effect on child-robot interaction. The researchers added a skin patch to a robot named KASPAR and proposed a new tactile recognition algorithm. The proposed approach allows automatic identification of touch events with high accuracy, i.e. 84 out of 100 cases of human touching the robot has been correctly detected. The system can be used to deal with both hypersensitivity, i.e. over sensitivity, and hyposensitivity, i.e. under sensitivity, to touch of children with autism.

### 3.2.2 Inflexible Adherence to Routines sub-category

Software sample:

*Astrojumper: motivating children with autism to exercise using a VR game* [78]. The authors believe that motivating children with autism to exercise is often hard, since they usually stick to few specific routines and interests in their lifestyle. One of these interests is their desire to play video games. Therefore, the authors utilized this interest and designed a virtual reality game in order to motivate these children to exercise. Their preliminary results show that a group of children with autism were very interested in the game.

Robots Sample:

*Reversal Learning Task in Children with Autism Spectrum Disorder: A Robot-Based Approach* [79]. Costescu et al. [79] think robots are capable of reducing the inflexible behavior of children with autism. Therefore, they proposed the use of Keepon robot in a cognitive flexibility task. However, their results show that children with autism had poorer performance when the robot interfere the task. This could be because of the fact that the robot might distract children from their main task. They believe that more study is needed in order to validate or invalidate these results.

### 3.2.3 Stereotyped or Repetitive Behaviors sub-category

Software sample:

*Behavior Imaging: Using Computer Vision to Study Autism* [71]. Rehg [71] believes that machine vision approaches can be utilized to recognize and analyze specific behaviors of children with autism. Focus of this research is mainly on recognizing social interaction from video streams. Hence, they proposed new methods based on quasi-periodic pattern analysis.

Devices Sample:

*Recognizing Stereotypical Motor Movements in the Laboratory and Classroom: A Case Study with Children on the Autism Spectrum* [72]. Albinali et al. [72] designed a system which automatically recognizes stereotypical behaviors, body rocking, and hand flapping in children with ASD. They believe that using wearable accelerometers alongside utilizing feature extraction and classification methods are suitable for recognizing activities such as postures, ambulation, exercise, and household activities.

Their data collection approach included six participants with autism. For each participant, they studied body rocking, hand flapping, and/or both in both laboratory, i.e. a controlled environment, and a school, i.e. uncontrolled/semi-controlled environment. Each person had to wear three wireless sensors, two of them on both wrists and one on the torso. The sensors transmitted 3-axis +/- 2g motion at 60HZ to a receiver which send data to a computer where data were synchronized and stored. In parallel, they captured a video in order to annotate each activity. They extracted the following five features:
- The distances between the means of the axes of each accelerometer to capture sensor orientation for posture,
- Variance to capture the variability in different directions,

- Correlation coefficients to capture the simultaneous motion in each axis direction,
- Entropy to capture the type of stereotypical motor movement, and
- FFT peaks and frequencies to capture differentiation between different intensities of the stereotypical motor movements.

In the laboratory setting, participants were asked to sit on a comfortable chair. Then their teacher asked them to bring objects like books and toys. In addition, in school setting, participants were observed in typical activities, such as eating lunch, spelling program, and sorting items, in two separate situations, i.e. doing these activities alone and with their teacher.

### 3.3 Adversities in Learning category

Software-based approaches in this category addressed both Difficulties in Daily Living Skills and Difficulties in Academic Skills sub-categories. In detail, there are studies that focused on teaching daily living needs such as useful vocabularies [18], [22], [24], [80], enhancing speech problems [18], and teaching essential cognitive concepts [81]. For instance, Venkatesh et al. [81] designed an educational iPad game that teaches children basic concepts needed to accomplish daily tasks.

The majority of studies in developing software-based systems to teach academic skills are focuses on teaching language skills [82]–[85]. Among the limited number of studies in other areas than language skills can name the work by Burton et al. [86] who used Video Self-Modeling (VSM) on an iPad for adolescent with autism to investigate how it affects the mathematics skill acquisition. Also, smith et al. [19] tried to teach basic science topics to children with autism. Kilroe et al. [87] developed a game that teaches number of relations, e.g. opposite/same, more/less, by using various visual stimuli to children with autism. Finally, a study investigated how technology can be helpful in classroom setting [88].

What comes next is a brief review of studies that used software-based approach to tackle Adversities in Learning sub-categories. It should be noted that we could not find any work using devices or robots to directly deal with adversaries in learning.

#### 3.3.1 Difficulties in Daily Living Skills sub-category
  Software sample:

*Acquisition and Generalization of Chained Tasks Taught with Computer Based Video Instruction to Children with Autism* [89]. In this research, Ayes et al. [89] analyzed the effect of a computer software that teaches daily living skills to children with autism through video based instructions. In detail, the focused on skills, such as setting a table, making soup, and make a sandwich. The tasks are then divided into small steps. Children should follow these steps respectively to accomplish the skills. The researchers utilized the ABA analysis technique used by therapist working with children with autism.

#### 3.3.2 Difficulties in Academic Skills sub-category
  Software sample:

*vSked: Evaluation of a System to Support Classroom Activities for Children with Autism* [90]. Hirano et al. assumed that using visual symbols is a promising approach to involve children with special needs in understanding the structure of activities in their lives. Thus they designed and developed an application which provides user friendly tools and setup for visual schedules display while eliminating the hassle of dealing with cards and boards. In the traditional paper-based symbol scheduling, the teachers or parents have to maintain a large amount of small cards and provide materials for future plans. In contrast, vSked allows the users to customize their application, through the ability to add or remove pictures or create a desired activity. Moreover, their system provides an automatically reporting mechanism which clearly diminishes the need of spending a lot of time to capture data and analyze them. In addition, their system consists of a large touch screen display, a personal screen for teachers which provides an administrator control center, and hand-held touch screen devices for each participant in a classroom. This structure obviously delivers a good communication and information sharing system among all the students.

## 4 Discussion

It can be easily concluded that the most popular type of technology that researchers used for screening, assessment, and rehabilitation of children with autism is software-based systems. Accessibility, affordability, ease of design, development, test, and use are definitely the main reasons for the wide spread use of software-based systems. Also, according to Fig. 4, the use of technology in screening, assessment, and rehabilitation of children with autism are mainly focused on "Difficulties in Social and Communication Skills" category. The wide range of studies in this category could be because of its role in effective screening and assessment and being the main focus in rehabilitation of children with autism. Furthermore, it is easier to develop technologies for social and communication skills than the repetitive behaviors. Obviously, dedicated devices are not as suitable as robots and software-based systems in this category. This is due to the fact that a suitable approach should be based on human-computer or human-robot interactions.

Moreover, as it can be inferred from Fig. 4, there exist few studies in "Restricted or Repetitive Behaviors" category. The reason behind this could be its lower importance in comparison to other categories. In fact, utilizing technologies in order to enhance "Difficulties in Social and Communication Skills" category is shown to be effective and practical. Based on this, it is hopeful that the related technologies might be implicitly effective in "Restricted or Repetitive Behaviors" category. In detail, when a proposed software-based technology is to rehabilitate a social and communication skill of children with autism, it might keep away the children from their repetitive behaviors.

Despite the current lack of studies using dedicated devices and robots in "Adversities in Learning", it is expected to see more devices and robots used for this purpose. The reason is that relevant devices and robots are becoming cheaper and widely available [91]. Furthermore, there are studies showing the effect of using robots in teaching normal children [92], which may be extended into children with autism.

A more detailed analysis on the data (Fig. 5) shows that although no study has been found in the "Fixated Interests" sub-category, studies based on activity recognition methods [93], [94] can be proposed in this sub-category. Specific wearable devices like E4 wrist band [16], can be utilized in the "Hyperactivity to Sensory Input" sub-category. In addition, robots could probably provide more effective solutions to "Inflexible Adherence to Routines". In fact, the interest of children with autism in interaction with robots can be employed in order to motivate children to change their extreme adherence to specific activities. Additionally, dedicated devices can be used to detect "Stereotyped or Repetitive Behaviors" using pattern recognition algorithms. Also, robots and software-based approaches can be employed in detecting stereotyped or repetitive speech of children with autism.

According to Fig. 6, the type of used technology in "Difficulties in Social and Communication Skills" category, relatively depends on its symptoms. For instance, several previous studies preferred to propose software-based approaches in tackling eye contact difficulties, rather than utilizing robots. On the other hand, there are few researchers studied "No/Few Understanding Body Orientation", using software-based systems. This could be due to the embodiments of robots which makes them more suitable to teach understanding gestures and performing body language. As a general rule of thumb, robots are suitable for tasks that need 3D environment setup and software-based approaches are suitable for tasks that can be done in 2D environment too. It also should be noted that implementing special cases, such as teaching emotions in different scenarios, are easier in software-based systems rather than using robots.

## 5 CONCLUSION

In this paper, a comprehensive review of utilized technologies in screening, assessment, and rehabilitation of children with autism has been presented. The focuses of previous related surveys were limited to specific type(s) of technology or limited to few number of ASD symptoms. Hence, in this study, the types of the utilized technologies divided into two general parts, i.e. software-based and hardware-based, which by itself where divided into dedicated devices and robots. Then, technology types of 212 previous studies were categorized into three main areas of research in autism screening, assessment, and rehabilitation. Two of these three categories have been determined based on the ASD diagnostic criteria provided in Diagnostic and Statistical Manual of Mental Disorders (DSM-V) [1]. The third category, i.e. adversities in learning, is designed to include all the studies related to teaching daily life knowledge and skills. Also, each category has been divided into sub-categories to provide better evaluation and analysis of the previous studies. This categorization and sub-categorization allowed us to determine unexplored or barely explored areas which can be a guide for researchers to plan their future research.

Finally, in each category and sub-category, the general trend has been presented. Then, for each sub-category sample studies, one for each one of three technologies, are presented in more detail to better demonstrated the research activities in a given sub-category.

As a result, newcomer in the field can easily look through these statistics to observe what type of technology is more promising for each symptom. Also, the common intervention procedures for any symptoms can be found by reviewing sample studies.